\documentclass[12pt]{article}

\usepackage{fullpage}
\usepackage{amssymb, amsmath, bm, bbm, nicefrac}
\usepackage{graphicx, subcaption, booktabs, enumitem}
\usepackage{setspace, lineno}
\usepackage[authoryear]{natbib}

\allowdisplaybreaks  % Allow page breaks within align
\providecommand{\keywords}[1]{\textbf{Keywords:} #1}

\newcommand{\bmu}{\bm{\mu}}
\newcommand{\balpha}{\bm{\alpha}}
\newcommand{\bbeta}{\bm{\beta}}
\newcommand{\bSigma}{\bm{\Sigma}}
\newcommand{\bPsi}{\bm{\Psi}}
\newcommand{\bX}{\mbox{\bf X}}

\newcommand{\bZ}{\mbox{\bf Z}}
\newcommand{\bY}{\mbox{\bf Y}}
\newcommand{\bT}{\mbox{\bf T}}

\newcommand{\bs}{\mbox{\bf s}}
\newcommand{\iid}{\stackrel{\mathrm{iid}}{\sim}}
\newcommand{\indep}{\stackrel{\mathrm{indep}}{\sim}}
\newcommand{\calS}{{\cal S}}
\newcommand{\Matern}{\mbox{Mat$\acute{\mbox{e}}$rn}}
\newcommand{\KL}{\mbox{Karhunen-Lo$\acute{\mbox{e}}$ve}}
\newcommand{\beq}{\begin{linenomath} \begin{equation}}
\newcommand{\eeq}{\end{equation} \end{linenomath}}

\title{A nonparametric spatial test to identify factors that shape a microbiome} 
\author{Susheela P. Singh\thanks{North Carolina State University, Raleigh, NC} , Ana-Maria Staicu\footnotemark[1] , Robert R. Dunn\footnotemark[1] , \\ Noah Fierer\thanks{University of Colorado, Boulder, CO} and Brian J. Reich\footnotemark[1]}

\begin{document}

\maketitle

\begin{abstract}
The advent of high-throughput sequencing technologies has made data from DNA material readily available, leading to a surge of microbiome-related research establishing links between markers of microbiome health and specific outcomes.  However, to harness the power of microbial communities we must understand not only how they affect us, but also how they can be influenced to improve outcomes. This area has been dominated by methods that reduce community composition to summary metrics, which can fail to fully exploit the complexity of community data. Recently, methods have been developed to model the abundance of taxa in a community, but they can be computationally intensive and do not account for spatial effects underlying microbial settlement. These spatial effects are particularly relevant in the microbiome setting because we expect communities that are close together to be more similar than those that are far apart. In this paper, we propose a flexible Bayesian spike-and-slab variable selection model for presence-absence indicators that accounts for spatial dependence and cross-dependence between taxa while reducing dimensionality in both directions. We show by simulation that in the presence of spatial dependence, popular distance-based hypothesis testing methods fail to preserve their advertised size, and the proposed method improves variable selection. Finally, we present an application of our method to an indoor fungal community found with homes across the contiguous United States. \\

\noindent \keywords{Bayesian nonparametrics, Dirichlet process, high dimensional data, spatial modeling, spike-and-slab prior, variable selection}
\end{abstract}

\newpage
\section{Introduction}\label{s:intro}
The development and increased accessibility of high-throughput sequencing technologies have steadily decreased the cost of studying DNA \citep{Reuter:2015, Heather:2016}.  This has made analysis of microbial communities found in environmental samples easier. Armed with previously cost-prohibitive data, investigators have published a flurry of work leveraging microbiome information with applications in varied fields including forensics, ecology, archeology, and public health. To date, much of this work has focused on studying abiotic and biotic factors that structure microbial communities and on identifying links between microbiome characteristics (e.g., composition or diversity) with specific outcomes. For example, studies have shown that microbiome composition can identify the source of a sample \citep{Grantham:2015}, linked changes in the gut microbiome to immune system dysfunction \citep{Round:2009}, tied reduced microbial diversity to obesity \citep{Turnbaugh:2009}, and connected imbalances in composition to Type 2 diabetes \citep{Qin:2012}. Though there has been an increased focus on defining the characteristics and markers of ``healthy'' microbiome communities for various systems within the body \citep{Human:2012, Ravel:2011}, the tools to understand which factors may exert influence on microbiome composition are limited. 

In this paper, we consider data from \citet{Barberan:2015}, which contains presence-absence indicators for over 57,000 fungal taxa based on dust samples from 1,331 homes in the contiguous United States. In addition, we have geographic, climatic, and household covariate information at each sampling location covering a wide range of explanatory variables. Our objective is to develop a testing procedure to identify covariates that influence microbiome composition that is applicable to high-dimensional, spatial, binary data and leverages the multivariate dependence between microorganisms.

Previous studies have demonstrated that a home's location, design, its occupants, and their activities, can all influence the microbiome composition present in dust within the home \citep{Barberan:2015, Kettleson:2015, Dannemiller:2016}.  These studies generally reduce the data to summary measures (e.g., richness, Shannon Diversity index) or a measurement of dissimilarity in composition between samples such as Bray-Curtis dissimilarity \citep{Bray:1957}. Often, investigators then test for association between environmental covariates and these summaries using nonparametric permutation-based tests, the most popular of which are ``ANalysis Of SIMilarities'' \citep[ANOSIM;][]{Clarke:1993} and ``PERmutational Multivariate ANalysis Of VAriance'' \citep[PERMANOVA;][]{Anderson:2001, McArdle:2001}. A tenuous assumption of these tests is exchangeability across sampling locations; we show that violation of this assumption inflates Type I error rates. This is of particular importance in our motivating example because \citet{Barberan:2015} note that nearby sampling locations exhibit more similar fungal communities than those that are far apart, and thus the assumption of exchangeability is known to be violated.

Distance-based methods are also limited in interpretability.  Because they partition the pairwise distances between samples, we cannot determine precisely how a covariate affects the composition or which taxa are directly affected. In a setting where an investigator may endeavor to target an intervention at a specific taxon or group of taxa, these tests are insufficient. Techniques such as redundancy analysis and canonical correspondence analysis are commonly used tools that can allow these relationships to be specified, but they too rely on permutation-based tests with an underlying assumption of independence across sampling locations.  Recently, methods addressing similar concerns have been developed for use on the compositional taxa counts \citep{Chen:2013, Zhao:2015, Grantham:2017, Wadsworth:2017, Wang:2017}. However, these methods are not appropriate for binary data and do not address spatial dependence in the data.  Additionally, the proposed methods in \citet{Chen:2013} and \citet{Wang:2017} rely on optimization routines that may not be suitable for problems with thousands of sample locations and tens of thousands of taxa. \citet{Grantham:2017} introduces a mixed effects model that accounts for correlation between taxa, but not between sampling locations. \citet{Warton:2011} proposes a permutation-based test that analyzes the community response and is applicable to presence-absence data, but it too relies on an assumption of spatial independence and is computationally expensive, and thus it is infeasible for large problems. \citet{Clark:2017} provides a framework to unify disparate data types, including presence-absence indicators, but it does not account for spatial dependence, does not incorporate dimension reduction, and does not perform variable selection or covariate testing.

As an alternative, we propose a flexible Bayesian variable selection method that uses a spike-and-slab prior and accounts for spatial dependence between nearby samples and cross-dependence between taxa. A unique feature of microbiome data is the large number of taxa, and we exploit this feature to estimate a nonstationary spatial covariance function using data-driven basis functions \citep{Lorenz:1956} and to relax the normality assumption common in spatial analysis \citep{Nelsen:1999, Gelfand:2005, Reich:2007, Petrone:2009, Rodriguez:2010}. \cite{Shirota:2017} proposes a nonparametric model for presence-absence data, but their aim is prediction rather than variable selection and testing for covariate effects. We provide a global test of whether or not environmental covariates affect microbiome composition that is interpretable, reliable, and has fully characterized uncertainty. In addition, our method produces clusters of taxa and tests for covariate effects on individual taxa.

The remainder of the paper is structured as follows: in Section \ref{s:data}, we further describe the data; in Section \ref{s:model}, we detail the modelling procedure; in Section \ref{s:NSBasis}, we propose a procedure to estimate data-driven basis functions; in Section \ref{s:sim}, we present a simulation study comparing our proposed method to several competitors; and in Section \ref{s:analysis}, we apply the proposed method to an indoor fungal community and compare our results to a previous study.  Finally, we conclude with a brief summary in Section \ref{s:con}.

\section{Motivating Data}\label{s:data}
Wild Life of Our Homes (WLOH; yourwildlife.org) is a citizen-science project focused on studying microbial diversity in and around our homes. As part of the project, participants received sampling kits and instructions specifying nine standardized locations around their homes at which samples should be taken \citep{Dunn:2013}. The returned swabs were prepared using the direct PCR approach \citep{Flores:2012}, which amplifies the DNA present in the samples and allows them to be sequenced and classified into Operational Taxonomic Units (OTUs). The total amount of genetic information in a sample is an artifact of the sequencing process, and as a result, the raw number of sequenced reads identified for a given OTU is not comparable across samples. Thus, rather than analyzing the read counts directly, we consider the presence-absence indicators for each taxon. This transformation to presence-absence does not entirely remove the effects of the sequencing process from the data.  For example, a sample with a low total number of reads may still incorrectly consider too many taxa as absent. However, the transformation tempers the effect in most other cases.

\begin{figure}[ht]
\centering
\begin{subfigure}{0.46\textwidth}
\centering
\includegraphics[width=\textwidth, clip, trim=1cm 2cm 0.75cm 1cm]{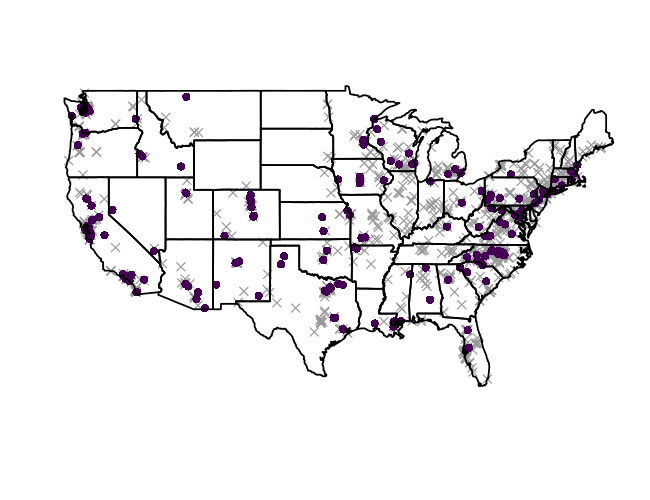}
\caption{\emph{Trichosporon asahii}}
\end{subfigure} \qquad
\begin{subfigure}{0.46\textwidth}
\centering
\includegraphics[width=\textwidth, clip, trim=1cm 2cm 0.75cm 1cm]{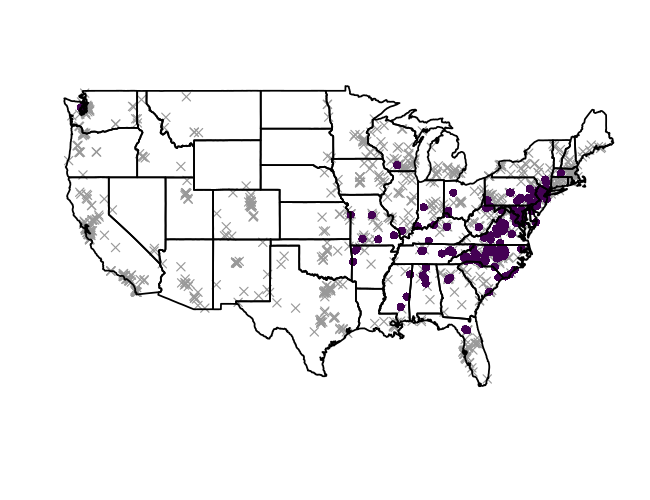}
\caption{\emph{Perenniporia narymica}}
\end{subfigure}
\caption{Map of presence (purple circle) or absence (gray $\times$) for two primarily indoor fungal taxa at each sampling location.}
\label{f:mapOTUs}
\end{figure}

In addition to supplying sample swabs, participants were asked to complete a questionnaire providing details about the home's location, design features, and its occupants. Geographic and climatic information were collected based on latitude and longitude from the Climate Research Unit Time Series v3.21 Dataset \citep{Harris:2014} and the National Land Cover Database \citep{Fry:2011} for a total of over $170$ covariates. From samples collected between 2012 and 2015, data was successfully sequenced for 1,331 homes spanning the 48 contiguous United States and the District of Columbia indicating the presence of 57,304 distinct fungal taxa.  Of these, we focus on $m=763$ taxa identified in \citet{Barberan:2015} as being more prevalent indoors than outdoors and on a set of $p=20$ potentially influential covariates similar to those in their analysis. The presence or absence at each sampling location for two of these taxa are mapped in Figure \ref{f:mapOTUs}. 
In the left panel, \emph{Trichosporon asahii}, which is commonly found living on human skin, is seen to be widespread while in the right panel, \emph{Perenniporia narymica} is seen to occur mainly in the mid-Atlantic region. Thus, there is evidence both that there is spatial dependence underlying the presence of fungal taxa and that the strength of that dependence varies across taxa.

\section{Nonparametric Spatial Model}\label{s:model}
Let $Y_{j}(\bs)$ be the binary indicator that OTU $j=1,\hdots,m$ is present in the sample at spatial location $\bs$. Suppose that we have a set of $p$ covariates, $\bX(\bs) = [X_{1}(\bs),\hdots,X_{p}(\bs)]$, such as those described in Section \ref{s:data}. We assume there exists a latent continuous process $Z_{j}(\bs)$ such that $Y_{j}(\bs) = \mathbbm{1}\{Z_{j}(\bs)>0\}$. The latent process is modelled as
\begin{equation}\label{eq:latent}
   Z_{j}(\bs) = \beta_{0j} + \bX(\bs)\bbeta_{j} + e_{j}(\bs), 
\end{equation}
where $\beta_{0j}$ is an intercept and $\bbeta_{j} = (\beta_{j1}, \hdots, \beta_{jp})^{\prime}$ are regression coefficients that together model the probability that OTU $j$ is present in a particular location. The final term, $e_{j}(\bs)$, is a multivariate spatial process with $\mbox{E}[e_{j}(\bs)]=0$ and $\mbox{Var}[e_{j}(\bs)]=1$ that models dependence not captured in the covariates between spatial locations and between OTUs.  This defines a probit link for the binary responses, $\mbox{P}\left[Y_{j}(\bs) = 1 \, | \, \bX(\bs)\right] = \Phi\left[\beta_{0j} + \bX(\bs)\bbeta_{j}\right]$, where $\Phi$ is the standard normal cumulative density function. The assumption that $\mbox{Var}[e_{j}(\bs)]=1$ is necessary because the covariate magnitudes are identifiable only up to the ratio of effect size to variance.

Our primary goal is to develop a test to identify factors that influence microbiome composition. A covariate influences the composition if it affects the probability that \emph{any} of the taxa will be present in a location, and thus we test the global hypotheses
\begin{equation} \label{eq:hypothesis}
\mbox{H}_{0r}: \, \beta_{jr} = 0 \text{ for all } j \quad \text{versus} \quad \mbox{H}_{1r}: \, \beta_{jr} \neq 0 \text{ for some } j,
\end{equation}
where $r$ and $j$ denote the covariate and OTU indices, respectively. The structure of this global test provides a means to identify an influential factor even if it affects only a small subset of the OTUs. 

It remains to describe the modelling procedure for the individual components identified in \eqref{eq:latent}. In Section \ref{s:SSVS} we specify a Bayesian variable selection model for the regression coefficients, $\bbeta_{j}$, and in Section \ref{s:DPM} we specify a nonparametric Bayesian model for the multivariate spatial process, $e_{j}(\bs)$.

\subsection{Identifying influential covariates} \label{s:SSVS}
We use a spike-and-slab prior for the coefficients, $\beta_{jr}$, to perform variable selection \citep{Mitchell:1988, George:1993, Kuo:1998}. We assume that each coefficient can be written as $\beta_{jr} = \delta_{jr}\gamma_{jr}$ for an inclusion indicator, $\delta_{jr} \in \{0, 1\}$, and magnitude, $\gamma_{jr} \in \mathbb{R}$. This formulation allows us to simplify the hypotheses in \eqref{eq:hypothesis} in terms of the number of OTUs for which the $r\textsuperscript{th}$ covariate is included, $M_{r} = \sum_{j=1}^{m}\delta_{jr}$: 
\begin{equation}\label{eq:hypothesisM}
\mbox{H}_{0r}: \, M_{r} = 0 \quad \text{versus} \quad \mbox{H}_{1r}: \, M_{r} > 0.
\end{equation}
To evaluate this, we calculate the posterior probability of the null hypothesis, $\mbox{P}\left(M_{r}=0 \, | \, \bY \right)$, and compare to a threshold $t \in [0, 1]$. If the posterior probability of the null hypothesis is below the threshold, then the covariate is deemed influential.

Because we do not want to include the intercept in the variable selection process, we give it a separate prior $\beta_{0j} \iid \mbox{N}(0, \tau_{0}^{-1})$ with $\tau_{0} \sim \mbox{Gamma}(a_{0}, b_{0})$.  Similarly, the magnitudes have the standard conjugate formulation, $\gamma_{jr} \indep\mbox{N}(0, \tau^{-1}_{r})$ with $\tau_{r} \iid \mbox{Gamma}(a_{r}, b_{r})$. The inclusion indicators are distributed $\delta_{jr} \indep \mbox{Bernoulli}(\pi_{r})$, where $\pi_{r}$ is the prior inclusion probability for the associated covariate.

The prior on $\pi_{r}$ is chosen to induce sparsity in the coefficients such that the prior probability of the global null hypothesis in \eqref{eq:hypothesisM} is $0.5$, reflecting no prior knowledge of whether or not a covariate is influential. In particular, the inclusion probabilities have prior density
\begin{equation}\label{eq:p_inc}
\mbox{P}\left(\pi_{r}\right) = \omega\left[\frac{1}{\mbox{B}(1, \theta)}(1-\pi_{r})^{\theta-1}\right] + (1-\omega),
\end{equation}
a mixture of $\mbox{Beta}(1, \theta)$ and $\mbox{U}(0, 1)$ distributions weighted by $\omega \in [0,1]$ and with $\theta \ge 1$. This prior has large mass on the sparse model with $\pi_{r}$ near 0, as is common in high-dimensional Bayesian variable selection \citep{Castillo:2012, Zhou:2015, Rovckova:2016}, but remains flexible enough to allow substantial probability for large values of $\pi_{r}$. As $\omega$ approaches 1, the prior inclusion probabilities are driven toward 0, leading to sparser coefficient vectors as in the oft used $\mbox{Beta}(1, \theta)$ special case, and as $\omega$ decreases to 0 the uniform component dominates and covariates will be added more readily. We can also influence the level of sparsity in the coefficients through the parameter characterizing the Beta distribution, $\theta$. If $\theta=1$ then the prior is simply $\mbox{U}(0,1)$, and the coefficient vectors will not be sparse.  As $\theta$ increases, the density associated with large values of $\pi_{r}$ decays sharply, while density associated with small values changes less drastically, leading to a steeper density curve. As a reasonable default, fix $\omega=0.5$ and set $\theta=m^{2}$, where $m$ is the number of taxa under consideration, which gives $\mbox{P}\left(M_{r}=0\right) = 0.5$ \emph{a priori} for each covariate, as desired.

\subsection{Capturing residual dependence}\label{s:DPM}
As we show in Section \ref{s:sim}, properly accounting for residual dependence is necessary for valid statistical inference. To model the residual dependence in \eqref{eq:latent}, we assume that $e_{j}(\bs)$ can be decomposed into a structural component, $\xi_{j}(\bs)$, and an independent component (or nugget), $\epsilon_{j}(\bs)$, such that $e_{j}(\bs) = \xi_{j}(\bs) + \epsilon_{j}(\bs)$. The structural component contributes variance $\rho \in [0, 1]$, leaving the nugget distributed $\epsilon_{j}(\bs) \iid \mbox{N}(0, 1-\rho)$ to satisfy the identifiability constraint that $\mbox{Var}[e_{j}(\bs)] = 1$. We use a basis expansion model for $\xi_{j}(\bs)$ and write $\xi_{j}(\bs) = \bm{\Psi}(\bs)\balpha_{j}$, where $\bm{\Psi}(\bs) = [\psi_{1}(\bs),\hdots,\psi_{L}(\bs)]$ are orthogonal spatial basis functions common to all taxa and $\balpha_{j} = (\alpha_{j1},...,\alpha_{jL})^{\prime}$ are their associated loadings, for $L$ finite or infinity. The model for the process now becomes $e_{j}(\bs) = \bm{\Psi}(\bs)\balpha_{j} + \epsilon_{j}(\bs)$.

We use a Dirichlet process prior \citep{Ferguson:1973} for the distribution of the loadings, which can be written as $\balpha_{j} \iid f(\balpha)$, where $f$ is the infinite mixture
\beq\label{eq:DPprior}
   f(\balpha) = \sum_{k=1}^{\infty}p_k\mathbbm{1}\{\balpha = \bmu_{k}\}.
\eeq
The mixture means have priors $\bmu_{k} \iid \mbox{N}(\bmu_0, \rho \mbox{\bf{I}}_{L})$, where $\bmu_{0} \sim \mbox{N}(\bm{0}, \tau_{\mu_{0}}^{-1}\mbox{\bf{I}}_{L})$, $\rho \sim \mbox{U}(0, 1)$, and $\tau_{\mu_{0}} \sim \mbox{Gamma}(a_{\mu_{0}}, b_{\mu_{0}})$. The mixture probabilities, $p_k$, are modelled using the stick-breaking representation \citep{Sethuraman:1994} wherein $p_1 = V_1$, $p_k = V_k\prod_{u<k}(1-V_u)$ for $k>1$, and $V_u \iid \mbox{Beta}(1,D)$.  This ensures that $p_k>0$ for all $k$ and $\sum_{k=1}^{\infty}p_k=1$ almost surely.  Rather than fix the Dirichlet process precision parameter, we assign it an uninformative positive prior, $D \sim \mbox{Gamma}(a_{d}, b_{d})$. With this infinite mixture model, our prior for the distribution of the spatial random effects, $\xi_{j}(\bs)$, has large support in the class of spatial processes \citep{Gelfand:2005}. In practice, the infinite mixture model in \eqref{eq:DPprior} is truncated at $K$ terms for computational purposes.  That is, we assume $g_k\in\{1,...,K\}$ for $K \le m$ by setting $V_K=1$, giving $f(\balpha) = \sum_{k=1}^{K}p_k\mathbbm{1}\{\balpha = \bmu_{k}\}$.

The Dirichlet process prior can be viewed as a clustering model for the spatial loadings over the OTUs. If we let $g_j \in \{1,2,\dots\}$ denote the cluster label for OTU $j$, then the mixture probability, $p_k$, can be interpreted as $\mbox{P}(g_{j}=k)$, the probability that OTU $j$ will be assigned to cluster $k$. Then, given that OTU $j$ has been assigned to cluster $k$, its associated spatial loading vector is the group mean for that cluster, i.e., $\balpha_{j} \, | \, g_{j} = k \text{ is } \bmu_{k}$. In the microbiome setting, it is reasonable to believe that taxa exhibit different spatial patterns, as in Figure \ref{f:mapOTUs}, and that groups of taxa will behave similarly.  For example, one may expect that organisms with similar functions or that require the same nutrients might be found in close proximity to one another. This leads to a natural expectation of clustering in the spatial effects over the OTUs.

In combination with the assumptions from the previous section, the model for the latent process becomes
\begin{linenomath} \begin{align*} \label{eq:latent_sp}
Z_{j}(\bs) &= \beta_{0j} + \bX(\bs)\bbeta_{j} + \bPsi(\bs)\balpha_{j} + \epsilon_{j}(\bs) \nonumber \\
  &= \beta_{0j} + \sum\limits_{r=1}^{p}X_{r}(\bs)\delta_{jr}\gamma_{jr} + \sum\limits_{l=1}^{L} \psi_{l}(\bs)\alpha_{jl} + \epsilon_{j}(\bs),
\end{align*} \end{linenomath}
where $\bbeta_{j}$ captures the covariates' effect on the probability that OTU $j$ will be present at location $\bs$, $\bPsi(\bs)\balpha_{j}$ captures residual spatial trends, and $\epsilon_{j}(\bs)$ are independent errors.  The details of the full proposed model and its implementation, as well as a discussion of its properties, are contained in the supplemental article \citep{Singh:2018}. We also show in the supplement that the covariance structure induced by our model is nonstationary in general, and that the strength of the Dirichlet process clustering controls the dependence between OTUs.

\section{Estimating the spatial basis functions} \label{s:NSBasis}
The model detailed in Section \ref{s:DPM} requires the construction of a set of spatial basis functions, $\bPsi(\bs)$, that are orthogonal and capable of reflecting nonstationarity. While there are several approaches available to estimate spatial basis functions from binary data \citep[e.g.,][]{Lee:2010}, we follow ideas from functional principal component analysis for binary-valued functional data and estimate the basis functions as the eigenfunctions of an estimated covariance function of the spatial latent process \citep{Hall:2008, Serban:2013}.

Let $\calS = \{\bs_{1},\hdots,\bs_{n}\}$ be the set of spatial locations at which the binary $Y_{j}(\bs)$ are observed. Our goal is to construct an estimator of the covariance of the latent process, $Z_{j}(\bs)$. To do so, we follow the Taylor approximation technique of \citet{Hall:2008}. Let $\sigma(\bs, \bs^{\prime})$ be the covariance between $\bZ(\bs)$ and $\bZ(\bs^{\prime})$, which for $\bs \neq \bs^{\prime}$ is estimated as
\beq \label{eq:Hall}
\hat{\sigma}(\bs, \bs^{\prime}) = \frac{\hat{\vartheta}(\bs, \bs^{\prime})}{\phi\{\hat{\nu}(\bs)\} \, \phi\{\hat{\nu}(\bs^{\prime})\}},
\eeq
where $\phi(\cdot)$ is the standard normal density function. This is akin to equation (10) in Hall et al. (2008), where the numerator, $\vartheta(\bs, \bs^{\prime})$, represents $\mbox{Cov}[\bY(\bs), \bY(\bs^{\prime})]$, and the denominator acts as a scaling factor, with $\nu(\cdot)$ denoting the mean of the latent process.

However, the component estimators differ from Hall et al. (2008) because we cannot assume that the latent processes share a smooth mean process. In our setting, the mean process may differ across taxa or may be non-smooth due to its dependence on non-smooth covariates. We first obtain $\hat{\eta}_{j}(\bs)$, the predicted probability that $Y_{j}(\bs)=1$ from separate probit regressions of $\bY_{j}$ onto $\bX$ for each taxon. Then we smooth $m^{-1}\sum_{j=1}^{m}\hat{\eta}_{j}(\cdot)$ over 2-D space using a bivariate kernel smoother to obtain an ``average'' mean process $\bar{\eta}(\cdot)$, and let $\hat{\nu}(\cdot) = \Phi^{-1}\{\bar{\eta}(\cdot)\}$, where $\Phi^{-1}(\cdot)$ is the standard normal quantile function. In order to obtain the estimated covariance of $\bY(\bs)$ and $\bY(\bs^{\prime})$, we calculate $m^{-1}\sum_{j=1}^{m}[Y_{j}(\bs)Y_{j}(\bs^{\prime}) - \hat{\eta}_{j}(\bs) \, \hat{\eta}_{j}(\bs^{\prime})]$ and smooth these estimates using a four-dimensional kernel smoother.  The resulting smoothed estimates are collected as $\hat{\vartheta}(\bs, \bs^{\prime})$. As is typical in nonparametric statistics, the optimal bandwidths are chosen using generalized cross-validation \citep{Craven:1978, Friedman:2001}. 

Applying this procedure to the variances will result in biased estimates \citep{Hall:2008}. To remove this bias, we consider a modified estimator, $\hat{\sigma}(\bs, \bs)$, and use the intercept of the weighted linear model
\begin{linenomath} \begin{equation*}
\hat{\sigma}(\bs, \bs^{\prime}) = \beta_{0} + w(\bs, \bs^{\prime})\,\mathrm{d}(\bs, \bs^{\prime})\beta + \epsilon,
\end{equation*} \end{linenomath}
for $\bs \neq \bs^{\prime}$ and with weights $w(\bs, \bs^{\prime}) = \mathrm{exp}\left[-\frac{d(\bs, \bs^{\prime})}{d_{10}}\right]\mathbb{I}(d(\bs, \bs^{\prime}) \le d_{10})$, where $d_{10}$ is the distance between $\bs$ and its $10\textsuperscript{th}$ closest neighbor for some distance measure $d$. In our application, we use the great-circle distance in miles.

Let $\hat{\bSigma}$ be the initial estimate of the spatial covariance matrix with elements $\hat{\sigma}(\bs, \bs^{\prime})$. By construction, $\hat{\bSigma}$ is symmetric. However, to ensure that it is positive semidefinite, we consider its low rank approximation. Let ${\tilde\phi}_1(\bs), \hdots, {\tilde\phi}_L(\bs)$ be the leading $L$ eigenvectors of $\hat{\bSigma}$, scaled by the square-root of their associated eigenvalues, such that they account for a specified percentage of explained variance. In our application, we use 90\%. To preserve the variance structure described in Section \ref{s:DPM} (i.e., $\mbox{Var}[\xi_{j}(\bs)] = \rho$), we need to ensure that $\sum_{l=1}^{L}\tilde{\phi}_{l}^{2}(\bs) = 1$.  If $L < n$, this will require scaling the eigenvectors to obtain
\begin{linenomath}\begin{equation*}
\psi_{l}(\bs) = \left[\frac{1}{\sum_{l=1}^{L} \tilde{\phi}_{l}^{2}(\bs)}\right]^{\frac{1}{2}}\tilde{\phi_{l}}(\bs).
\end{equation*} \end{linenomath}
Let $\bPsi = \left[\bm{\psi}_{1}, \hdots, \bm{\psi}_{L}\right]$, where $\bm{\psi}_{l} = \{\psi_{l}(\bs_{1}), \hdots, \psi_{l}(\bs_{n})\}^{\prime}$ for $l=1,\hdots,L$. After this scaling process, $\bPsi$ is no longer orthogonal on $\mathbb{R}^{L}$, and thus we rotate by its right singular vectors to obtain the proposed basis functions.  

Now, $\bPsi$ is scaled appropriately to preserve the variance structure we require, rotated to preserve orthogonality between basis functions, and reflects the nonstationarity we expect in the data. The estimated basis functions are available only at the locations in $\mathcal{S}$, and extrapolation would be required to make spatial predictions beyond the $n$ sample locations.  However, our objective is not spatial prediction, but rather to account for the complex dependence structure at the sampling locations to give a valid global test of covariate effects.

Because of the reliance on generalized cross-validation to select the bandwidth parameter, the four-dimensional smoothing step to obtain the $\hat{\vartheta}(\bs, \bs^{\prime})$ estimates can be prohibitively expensive. Two approaches to alleviating this burden are either to use a different method to select the bandwidth or to make the cross-validation less computationally intensive. As an example, a reasonable approach that avoids cross-validation might be to construct a variogram, identify the distance at which the correlation decays, and use that distance to set a bandwidth. Alternatively, if the data contains sampling locations that are close to one another, one could downsample the locations while approximately preserving the spatial coverage of the data. Then, generalized cross-validation can be done quickly on this smaller, representative set of locations to obtain an estimated optimal bandwidth. This latter approach is utilized in our data application in Section~\ref{s:analysis}.

\section{Simulation study} \label{s:sim}
In this study, we consider generating data while varying the type of spatial dependence in the latent process, the existence of cross-dependence between OTUs in the latent process, the magnitude of covariate effect size, and the degree of prevalence in covariate effects, and evaluate how these factors influence the true and false positive rates of the global test in \eqref{eq:hypothesisM}.

\subsection{Methods} \label{s:sim:design}
We generate data on a $15 \times 15$ grid on the unit square for a total of $n=225$ spatial locations.  For each of $m=50$ OTUs, we draw the latent process as $\bZ_{j} \sim \mathcal{N}_{n}\left(\bX\bbeta_{j}, 0.95\bSigma_{z} + 0.05\mbox{\bf{I}}_{n}\right)$. The structure of $\bSigma_{z}$ varies based on the type of spatial dependence:
\begin{enumerate}[leftmargin=*, labelindent=3.25em]
\item[(Ind)] Independence: $\bSigma_{z}=\mbox{\bf{I}}_{n}$,
\item[(Exp)] Stationary dependence: $\bSigma_{z}$ is populated by the exponential covariance function with spatial range set such that the correlation between the two closest sites is 0.75, and 
\item[(Nonstat)] Nonstationary dependence: where $\bSigma_{z}(\bs, \bs^{\prime}) = \mbox{cos}(2\pi s_{1})\mbox{cos}(2\pi s^{\prime}_{1}) + \mbox{sin}(2\pi s_{2})\mbox{sin}(2\pi s^{\prime}_{2})$ for $\bs = (s_{1}, s_{2})$.
\end{enumerate}
When the setting calls for multivariate dependence in the latent process, we assume a separable covariance function and define $\mbox{Cov}[Z_{j}(\bs), Z_{j^{\prime}}(\bs^{\prime})] = c(j, j^{\prime})\bSigma_{z}(\bs, \bs^{\prime})$, where $c(j, j^{\prime}) = 0.8^{|j-j^{\prime}|}$ is the cross-dependence function. In reality, we do not expect a meaningful ordering of the OTUs, but this covariance is used to generate data with a reasonable range of cross-correlations. The $p=20$ covariates are drawn from a mean-zero Gaussian process with separable covariance function $\mbox{Cov}\left[X_{r}(\bs), X_{r^{\prime}}(\bs^{\prime})\right] = c(r, r^{\prime})\bSigma_{x}(\bs, \bs^{\prime})$ where $c(r, r^{\prime})$ is as above, and $\bSigma_{x}$ is the exponential covariance with spatial range set such that the correlation between the two closest sites is 0.5.

Of the covariates, $p_{0}=6$ are influential (i.e., $\beta_{jr}$ is non-zero for some $j$) and the remainder are unimportant for all OTUs (i.e., $\beta_{jr} = 0$ for all $j$). In order to examine the ability of the algorithm to detect covariate effects across prevalences and magnitudes, the influential covariates are split into 3 pairs.  The first pair affects all OTUs, the second pair affects a randomly selected 50\% of OTUs, and the final pair affects a randomly selected 10\% of OTUs. Within each pair of non-zero coefficients, the first covariate is assigned a large magnitude of $\beta_{jr}=0.5$, and the second is assigned a small magnitude of $\beta_{jr}=-0.25$. The randomization over taxa for prevalence is done independently so that any one OTU may have 2, 4, or 6 important covariates.

Under each of the simulation settings we generate $N=50$ replicate datasets and fit the proposed spatial nonparametric model and several competing models:                                                                                                                                                                                                                                                                              
\begin{enumerate}[leftmargin=*, labelindent=3.25em]
\item[(PERM)] PERMANOVA \citep{Anderson:2001, McArdle:2001}, a permutation-based hypothesis test as implemented in the \texttt{R} package \texttt{vegan 2.4-3} using Bray-Curtis dissimilarity.
\item[(NS)] Nonspatial variable selection model, i.e., $\rho = 0$.
\item[(Mat)] Parametric spatial model where $\bm{e}_j = [e_{j}(\bs_{1}), \hdots, e_{j}(\bs_{n})]^{\prime}$ from \eqref{eq:latent} is modelled using a $\Matern$ covariance function. The smoothness has prior $\kappa \sim \mbox{U}(0, 2)$ \citep{Stein:1999, Banerjee:2005}, and the range has prior $\mbox{log}(\zeta) \sim \mbox{N}(0, \sigma^{2}_{\zeta})$ where $\sigma^{2}_{\zeta}$ is set such that the $99\textsuperscript{th}$ percentile of the prior distribution for the range is the maximum observed distance.
\item[(SNP)] Proposed nonparametric spatial model using the nonstationary basis detailed in Section \ref{s:NSBasis}, with the maximum number of groups set to $K=m$.
\end{enumerate}

For each of the Bayesian models (NS, Mat, and SNP), we fit the model using a special case of \eqref{eq:p_inc} where $\omega=1$ and $\theta=m$, which simplifies the prior to $\pi_{r} \iid \mbox{Beta}(1, m)$.  This commonly used prior on the inclusion probabilities will make it more likely for $\pi_{r}$ to be close to 0 than in the mixture setting. Our focus is on identifying covariates that are borderline cases, e.g., factors that influence only a few taxa. The sharper cut of this simplified prior near the origin makes the sampler less likely to include these covariate spuriously. To determine sensitivity to this prior specification, we also ran the simulation using the mixture prior in \eqref{eq:p_inc} with the recommended default values.  The results are qualitatively the same, with improved performance for Mat in identifying small magnitude covariates but a reduced ability to identify low prevalence covariates. The model performance for SNP is broadly unchanged. The remainder of the prior specifications are detailed in the supplemental article. The models are run for a total of 40,000 iterations with a burn-in period of 10,000, and the posterior samples are thinned by 2. We deem the $r\textsuperscript{th}$ covariate to be influential if the associated posterior probability of the null is below 0.05, i.e., $\mbox{P}\left(M_{r}=0 \, | \, \bY \right) < 0.05$, for the Bayesian models, or if its P-value from PERMANOVA is below 0.05. 

For each dataset, we evaluate the models using true positive rate (TPR) and false positive rate (FPR), presented in Table \ref{tab:sim}. Let $M^{*}_{r}$ be the indicator that the $r\textsuperscript{th}$ covariate is truly influential. The true positive rate is the percent of truly influential covariates correctly classified as influential by the model for a given threshold $t$, \[\mbox{TPR}(t) = \frac{\sum\limits_{r=1}^{p} M^{*}_{r}\mathbbm{1}\{\mbox{P}\left(M_{r}=0 \, | \, \bY \right) < t \}}{p_{0}}. \] The false positive rate is the percent of covariates classified by the model as influential that are not truly influential, \[\mbox{FPR}(t) = \frac{\sum\limits_{r=1}^{p} (1 - M^{*}_{r})\mathbbm{1}\{\mbox{P}\left(M_{r}=0 \, | \, \bY \right) < t \}}{p-p_{0}}. \] 
We also consider a ``registered'' true positive rate, where the threshold for each method is set to control its false positive rate at or below 0.05. In other words, for each model and simulated data set, we find the largest threshold $T$ such that $\mbox{FPR}(T) \le 0.05$, and use this calibrated threshold to evaluate the model.  In the case of PERMANOVA, the posterior probability of the null is replaced by the P-value. This allows us to compare the power of the methods on an even footing in Table \ref{tab:sim_reg}. Finally, in Table \ref{tab:sim_split}, we consider the inclusion rate for the influential covariates for each model, broken out by magnitude of the covariate effect, small (S) or large (L), and the prevalence of the covariate effect, 100\%, 50\%, or 10\%. The inclusion rate (IR) is defined as the proportion of the $N$ simulation runs for which the method correctly classified the covariate as influential, \[\mbox{IR}(t) = \frac{1}{N}\sum\limits_{s=1}^{N} \mathbbm{1}\{\mbox{P}\left(M_{s,r^{*}}=0 \, | \, \bY \right) < t \},\]
for each of the $r^{*}=1, \hdots, p_{0}$ influential covariates. As in the global results presented in Table \ref{tab:sim}, we use a fixed threshold of $t = 0.05$.

\subsection{Results}\label{s:sim:results}
\begin{table*}[h]
\caption{Summary of true positive rate (TPR), false positive rate (FPR), and average model fitting time in minutes for PERMANOVA (PERM), the nonspatial (NS), parametric $\Matern$ (Mat), and proposed nonparametric (SNP) models.}
\label{tab:sim}
\centering \footnotesize
\begin{tabular}{l|l||c|c|c||c|c|c}
& & \multicolumn{6}{c}{Dependence Between Taxa} \\
& & \multicolumn{3}{c||}{Independence} & \multicolumn{3}{c}{Autoregressive} \\
\hline
Spatial Dependence & Model & TPR & FPR & Time & TPR & FPR & Time \\
\hline
Independence & PERM & 0.62 & 0.05 & 3.75 & 0.49 & 0.06 & 3.68 \\
 & NS & 0.38 & 0.00 & 21.59 & 0.38 & 0.01 & 21.58 \\
 & Mat & 0.40 & 0.00 & 426.09 & 0.39 & 0.01 & 418.41 \\
 & SNP & 0.37 & 0.00 & 34.90 & 0.38 & 0.00 & 35.05 \\
\hline
Exponential & PERM & 0.96 & 0.80 & 3.98 & 0.87 & 0.61 & 3.75 \\
 & NS & 0.86 & 0.48 & 22.18 & 0.81 & 0.43 & 21.73 \\
 & Mat & 0.54 & 0.04 & 232.48 & 0.51 & 0.04 & 228.38 \\
 & SNP & 0.71 & 0.10 & 36.82 & 0.67 & 0.10 & 35.61 \\
\hline
Nonstationary & PERM & 0.81 & 0.49 & 3.74 & 0.81 & 0.49 & 3.88 \\
 & NS & 0.91 & 0.43 & 24.01 & 0.90 & 0.47 & 25.27 \\
 & Mat & 0.85 & 0.00 & 231.90 & 0.84 & 0.01 & 237.80 \\
 & SNP & 0.93 & 0.02 & 39.36 & 0.94 & 0.05 & 40.51 
 \\
\end{tabular}
\end{table*}

\begin{table*}[h]
\caption{Summary of ``registered'' true positive rate (TPR) and false positive rate (FPR) for PERMANOVA (PERM), the nonspatial (NS), parametric $\Matern$ (Mat), and proposed nonparametric (SNP) models. If values are not provided, there is no threshold value or significance level that controls the false positive rate at the required level.}
\label{tab:sim_reg}
\centering \footnotesize
\begin{tabular}{l|l||c|c||c|c}
& & \multicolumn{4}{c}{Dependence Between Taxa} \\
& & \multicolumn{2}{c||}{Independence} & \multicolumn{2}{c}{Autoregressive}\\ 
\hline
Spatial Dependence & Model & TPR & FPR & TPR & FPR \\
\hline
Independent & PERM & 0.63 & 0.05 & 0.48 & 0.05 \\
  & NS & 0.70 & 0.05 & 0.59 & 0.05 \\
  & Mat & 0.70 & 0.05 & 0.59 & 0.05 \\
  & SNP & 0.66 & 0.05 & 0.57 & 0.05 \\
\hline
Exponential & Mat & 0.58 & 0.05 & 0.53 & 0.05 \\
 & SNP & 0.63 & 0.05 & 0.56 & 0.05 \\
\hline
Nonstationary & Mat & 0.96 & 0.05 & 0.93 & 0.05 \\
 & SNP & 0.95 & 0.05 & 0.93 & 0.05 \\
\end{tabular}
\end{table*}

As is evident in Table \ref{tab:sim}, in the case of no spatial dependence in the data, PERM outperforms the Bayesian models.  The Bayesian tests are more conservative, but after tuning the FPR to be 0.05 (Table \ref{tab:sim_reg}), they have comparable or increased true positive rates as compared to PERMANOVA. The false positive rate for PERM is well-controlled even in the face of multivariate dependence, which is reasonable given that the permutation is done at the sampling location level and thus the structure of any cross-dependence between taxa is preserved.

However, in the presence of spatial dependence, PERMANOVA fails to preserve the size of the hypothesis test and has false positive rates an order of magnitude higher than expected. This is perhaps not unexpected as the pseudo-F test is built on the assumption of exchangeability across sampling locations.  Blind application of these permutation-based methods in settings where spatial independence across sampling locations is not a reasonable assumption will result in misleading conclusions.

When the data are spatially dependent, NS and PERM have high true positive rates accompanied by high false positive rates, indicating that the models favor including all covariates rather than discriminating between important and unimportant factors. Therefore, we exclude these models in Table \ref{tab:sim_split}, where we present the inclusion rate for the influential covariates broken out by prevalence and magnitude for each of the models. As before, in the case of spatial independence, PERM outperforms the Bayesian models, which all perform similarly. However in the case of spatial dependence, breaking out the model performance in this way allows us to see the contrast between the Bayesian spatial models. In particular, we can see that SNP outperforms the parametric model in identifying covariates with low prevalence and/or small magnitudes, which is our primary focus. Under the exponential covariance structure, SNP picks up the low prevalence, small magnitude covariate 16-20\% of the time, whereas the parametric model selects it in only 0-4\% of the replications. Similarly, under the nonstationary covariance structure, the parametric model selects the covariate in only 20-28\% of replications, as opposed to the 60-66\% of replications for SNP.

\begin{table*}[h]
\caption{Inclusion rate for influential covariates for PERMANOVA (PERM), the nonspatial (NS), parametric $\Matern$ (Mat), and proposed nonparametric (SNP) models, broken out by covariate magnitude (S=Small, L=Large) and prevalence (100\%, 50\%, 10\%).}
\label{tab:sim_split}
\centering \footnotesize
\begin{tabular}{l|l|l||c|c|c|c|c|c}
\multicolumn{2}{c|}{Dependence} & & \multicolumn{6}{c}{Covariate Prevalence and Magnitude} \\[0.1cm]
 & Between & \\
Spatial & Taxa & Model & 100\% L & 100\% S & 50\% L & 50\% S & 10\% L & 10\% S \\
\hline
Ind & Ind & PERM & 1.00 & 0.38 & 1.00 & 0.62 & 0.60 & 0.14 \\ 
 & & NS & 1.00 & 0.16 & 0.84 & 0.02 & 0.28 & 0.00 \\ 
 & & Mat & 1.00 & 0.18 & 0.86 & 0.02 & 0.32 & 0.00 \\ 
 & & SNP & 1.00 & 0.06 & 0.78 & 0.02 & 0.34 & 0.00 \\ 
 & AR(0.8) & PERM & 0.92 & 0.26 & 0.98 & 0.26 & 0.42 & 0.12 \\ 
 & & NS & 1.00 & 0.14 & 0.76 & 0.08 & 0.32 & 0.00 \\ 
 & & Mat & 1.00 & 0.14 & 0.76 & 0.08 & 0.36 & 0.00 \\ 
 & & SNP & 1.00 & 0.10 & 0.78 & 0.06 & 0.34 & 0.00 \\
\hline
Exp & Ind & Mat & 1.00 & 0.46 & 0.94 & 0.22 & 0.62 & 0.00 \\ 
 & & SNP & 1.00 & 0.76 & 0.96 & 0.56 & 0.76 & 0.20 \\ 
 & AR(0.8) & Mat & 1.00 & 0.38 & 0.94 & 0.18 & 0.50 & 0.04 \\ 
 & & SNP & 1.00 & 0.64 & 1.00 & 0.48 & 0.74 & 0.16 \\
\hline
Nonstat & Ind & Mat & 1.00 & 1.00 & 1.00 & 0.92 & 0.92 & 0.28 \\ 
 & & SNP & 1.00 & 1.00 & 1.00 & 1.00 & 0.98 & 0.60 \\ 
 & AR(0.8) & Mat & 1.00 & 1.00 & 1.00 & 0.88 & 0.94 & 0.20 \\ 
 & & SNP & 1.00 & 1.00 & 1.00 & 0.96 & 1.00 & 0.66 \\ 
\end{tabular}
\end{table*}

In addition, the spatial parametric model takes  6-10$\times$ longer to fit than the other models on average, and this is a relatively small problem with only $225$ locations and $50$ taxa. Mat requires several inversions of an $n~\times~n$ matrix during each MCMC iteration and it is clear that this becomes computationally infeasible for problems much larger than this simulated setting.  The proposed nonparametric model reduces the dimensionality of the problem for both large numbers of observations and a large number of observed taxa without sacrificing its aptitude to discern influential covariates from unimportant ones.

\section{Data Analysis}\label{s:analysis}
In light of PERMANOVA's demonstrated failure to preserve the size of the hypothesis test in the face of spatial and multivariate dependence, we revisit the analysis of \citet{Barberan:2015} in which the authors determined which, if any, of a set of environmental and household covariates affect the indoor fungal community composition of homes. The covariates of interest included mean annual precipitation (MAP), mean annual temperature (MAT), net primary productivity (NPP), elevation, age of the home, number of bedrooms, number of inhabitants, female-to-male ratio of the home's inhabitants, smoking status, number of dogs/cats/birds, whether or not the home has a basement, and number of days with the windows open. Using PERMANOVA, they find that the effects of outdoor variables and geographic location are more pronounced than the household covariates, but note that the presence of a basement in the home, the age of the home, and the presence of a dog also affect the composition of the indoor fungal microbiome.

We follow the intuition of \citet{Barberan:2015} and compile a similar list of covariates. In addition to those listed above, we include an indicator that the land is designated as forested, an indicator that the home is a rental unit, and the type of home (single family detached, multi-family dwelling, mobile). We replace the number of days with the windows open with the type of ventilation (central air-conditioning, central heat, window air-conditioning). NPP was missing for 81 of the sampling locations, and when considering only indoor fungal taxa, an additional 24 sampling locations had no present taxa. These locations have been removed, leaving $n=$ 1,226 locations and $p=20$ covariates in the analysis. 

\begin{figure}[ht]
\centering
\begin{subfigure}{0.46\textwidth}
\centering
\includegraphics[width=\textwidth, clip, trim=1cm 0cm 0.5cm 1cm]{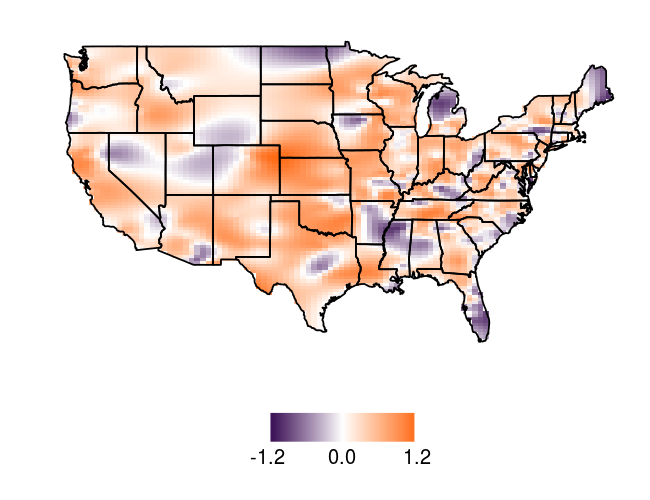}
\end{subfigure} \qquad
\begin{subfigure}{0.46\textwidth}
\centering
\includegraphics[width=\textwidth, clip, trim=1cm 0cm 0.5cm 1cm]{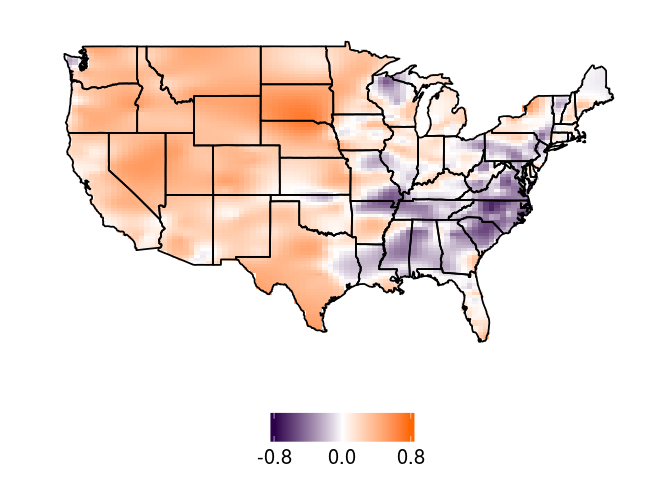}
\end{subfigure} \\
\begin{subfigure}{0.46\textwidth}
\centering
\includegraphics[width=\textwidth, clip, trim=1cm 0cm 0.5cm 1cm]{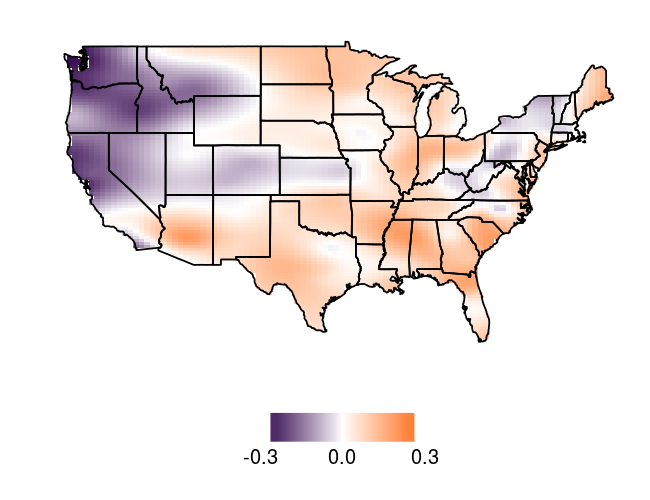}
\end{subfigure} \qquad
\begin{subfigure}{0.46\textwidth}
\centering
\includegraphics[width=\textwidth, clip, trim=1cm 0cm 0.5cm 1cm]{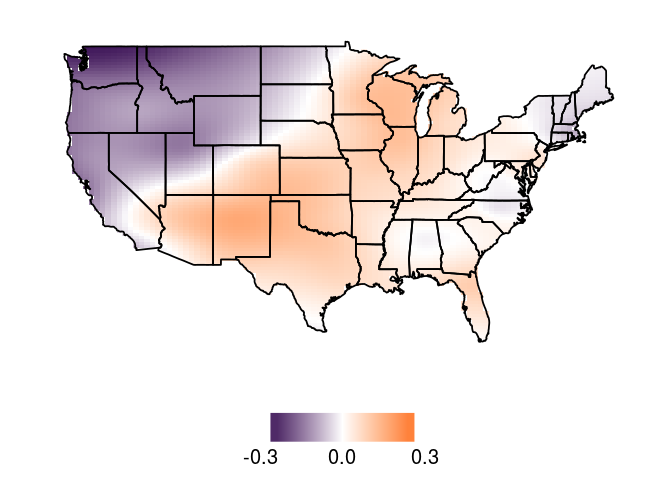}
\end{subfigure}
\caption{Maps of the first four spatial basis functions estimated from the WLOH data.}
\label{f:mapPsi}
\end{figure}

Using both PERMANOVA and the proposed nonparametric method, we investigated each covariate's ability to affect the composition of the taxa identified as the indoor fungal microbiome. SNP was run for 80,000 total iterations, keeping the final 30,000 posterior samples. Unlike in the simulation study, the maximum number of groups is set to $K=500 < m$. We utilized the downsampling strategy discussed in Section \ref{s:NSBasis} to build the spatial basis functions. The first few estimated basis functions are mapped in Figure \ref{f:mapPsi}.
The first several functions reflect the nonstationarity in the data, while later basis functions reflect smooth spatial variation. Reported in Table \ref{tab:data} for each covariate are the P-value from PERMANOVA, the posterior probability of the null hypothesis, the posterior expected number of taxa for which the covariate is selected, and a count of the number of taxa for which the associated coefficient value is positive or negative, assessed as $\sum_{j=1}^{763}\mathbbm{1}\{\mbox{P}(\beta_{jr} > 0 \, | \, \bY) > 0.975\}$ and $\sum_{j=1}^{763}\mathbbm{1}\{\mbox{P}(\beta_{jr} < 0 \, | \, \bY) > 0.975\}$, respectively, for the proposed model.

\begin{table}[ht]
\caption{Summary of variable selection results from PERMANOVA (PERM) and the proposed spatial nonparametric method (SNP). P-values are reported from PERM, and the posterior probability of the null hypothesis, the expected number of taxa for which the covariate is included, and the number of taxa for which the coefficient value is positive or negative are reported for SNP.}
\label{tab:data}
\centering \footnotesize
\begin{tabular}{l||c||c|c|c|c}
 & PERM & \multicolumn{4}{c}{SNP} \\
 \hline
Covariate & P-value & $\mbox{P}(\mbox{M}_{r} = 0 \, | \, \bY)$ & $\mbox{E}[\mbox{M}_{r} \, | \, \bY]$ & $\#$Positive & $\#$Negative \\
\hline
NPP & $<0.001$ & 0.00 & 445 & 38 & 161 \\ 
MAT & $<0.001$ & 0.00 & 349 & 40 & 122 \\ 
MAP & $<0.001$ & 0.00 & 131 & 5 & 14 \\ 
Central A/C & $<0.001$ & 0.00 & 117 & 6 & 5 \\ 
Multifamily dwelling & 0.038 & 0.00 & 82 & 9 & 0 \\ 
Forested & $<0.001$ & 0.00 & 35 & 0 & 0 \\ 
Elevation & $<0.001$ & 0.00 & 15 & 0 & 1 \\ 
Window A/C & $<0.001$ & 0.00 & 15 & 0 & 1 \\ 
Older home & 0.078 & 0.00 & 13 & 0 & 0 \\ 
Central heat & 0.015 & 0.03 & 11 & 0 & 0 \\ 
Basement & $<0.001$ & 0.04 & 8 & 0 & 0 \\ 
Number of dogs & 0.152 & 0.05 & 5 & 0 & 0 \\ 
Rental home & 0.075 & 0.16 & 4 & 0 & 0 \\ 
Number of occupants & 0.016 & 0.43 & 1 & 0 & 0 \\ 
Number of bedrooms & 0.386 & 0.46 & 1 & 0 & 0 \\ 
Mobile home & 0.289 & 0.48 & 1 & 0 & 0 \\ 
Smoking status & 0.756 & 0.49 & 1 & 0 & 0 \\ 
Percentage of females & 0.735 & 0.51 & 1 & 0 & 0 \\ 
Number of birds & 0.627 & 0.51 & 1 & 0 & 0 \\ 
Number of cats & 0.558 & 0.76 & 0 & 0 & 0 \\ 
\end{tabular}
\end{table}

Comparing the P-values from PERMANOVA and the posterior probability of the null hypothesis from SNP, we see that the two models largely agree, but we can identify several covariates that PERMANOVA includes at either the 0.05 or 0.10 significance level that would not be included in the SNP model. Given the inflated Type I error rates of the PERMANOVA test under spatial dependence in the simulation study, it seems likely that these are false positives. The proposed method is able to identify both covariates that are important to many taxa (e.g., MAT) and those that are important only to a few (e.g., whether or not a home is older). In addition, we are able to precisely describe \emph{how} covariates influence particular taxa.  For example, as one would expect, we note that most fungal taxa prefer cooler climes, but that there are some taxa that seem to thrive in the warmer temperatures. Generally, we corroborate the findings of \cite{Barberan:2015} and conclude that geographic and climatic factors are most influential to the indoor fungal microbiome composition. The household covariates that appear as influential are whether or not the home is older, the presence of a basement, 
whether or not the home is a multifamily dwelling, and whether or not the home has air-conditioning or central heating, all of which play a role in increasing the interaction between the indoor environment and the outdoors.

\begin{figure}[ht]
\centering
\begin{subfigure}{0.46\textwidth}
\centering
\includegraphics[width=\textwidth, clip, trim=1cm 2cm 0.75cm 1cm]{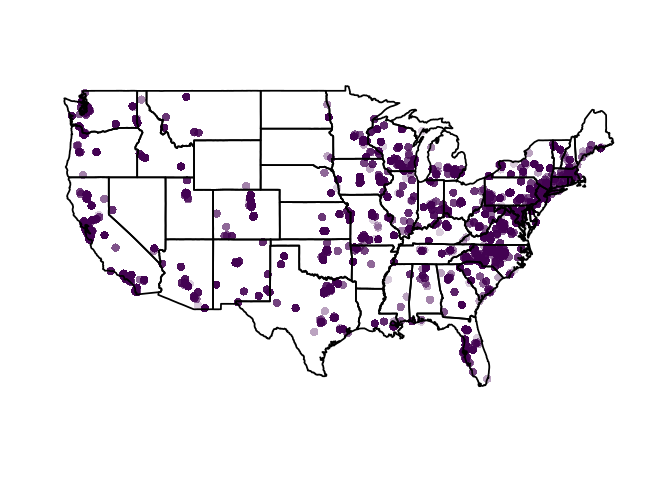}
\end{subfigure} \qquad
\begin{subfigure}{0.46\textwidth}
\centering
\includegraphics[width=\textwidth, clip, trim=1cm 2cm 0.75cm 1cm]{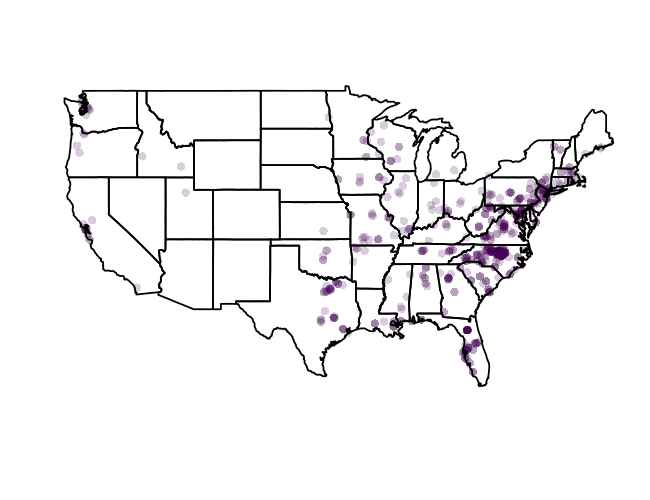}
\end{subfigure}
\caption{Map of presence for taxa assigned to a large cluster of 100 taxa and a small cluster of 3 taxa. A darker point indicates that a higher number of taxa are present in a location.}
\label{f:mapGroups}
\end{figure}

The 763 species are grouped into an estimated (posterior mean) 47 clusters. The largest clusters, based off of a $k$-means clustering algorithm with 47 clusters and using $1-\mbox{P}(g_{j}=g_{j^\prime})$ as the dissimilarity matrix, contain taxa that exhibit little spatial clustering and tend to be present across the country.  The smaller clusters tend to group together taxa that exhibit more localized presence.  For example, in Figure \ref{f:mapGroups}, the left panel displays the presence for the 100 taxa assigned to the largest cluster and the right panel displays the presence for the 3 taxa assigned to a smaller cluster.

In as much as our results add to those of previous analyses using data from the WLOH project, it is worth commenting about the additional biological insights our approach offers. \citet{Barberan:2015} found that, compared to bacteria, the composition of fungi in homes tended to be much more strongly driven by outdoor environmental conditions. In our analysis, this conclusion is even more strongly supported. The primary factors associated with differences in the composition of indoor fungi among households were those associated with climate and its effects, and nearly all (94.8\%) significant associations of individual taxa with particular covariates were associations with these environmental factors. 

Net Primary Productivity (NPP) was a particularly important correlate of the composition of indoor fungi. In the United States, NPP is highly correlated with forest cover, such that areas with higher NPP are almost always forests. In this light, it is perhaps not surprising that species more common in regions with high NPP were species associated with forests and dead and down wood, including multiple taxa of the species \emph{Xylobolus annosus}. Conversely, species that became less common under high NPP tended to be from the genera \emph{Alternaria}, \emph{Cladosporum}, \emph{Aspergillus}, and \emph{Phoma}, many of which are associated with decaying plant material. Fungi from decaying plant material, much of which is in leaf litter, might be more likely to become airborne in open habitats such as grasslands. Many species were also influenced by the direct effects of the mean annual temperature or precipitation in the region in which a house was located.

One of the few non-environmental covariates identified as influential was whether or not the home is a multifamily dwelling. Multifamily dwellings tended to favor fungi associated with human bodies or foods. These included three \emph{Candida} taxa, \emph{Cryptococcus oeirensis}, \emph{Penicillium concetricum}, and the brewer’s yeast (\emph{Saccharomyces cerevisae}). Also more common in these homes were \emph{Rhodotorula mucilagnosa} and \emph{Cystofilobasidium capitatum}, both of which do well under stressful conditions, such as those associated with bathrooms that are frequently cleaned. The way in which a house was heated or cooled also influenced which species were present. In particular, as has been noted in smaller scale studies \citep{Hamada:2002}, we confirm here that houses with air conditioning tend to be more likely to have \emph{Cladosporium} and \emph{Penicillium} fungi, which are known to grow in air conditioning units and then spread through houses. Air conditioners were also associated with several other fungal species, including the wood rot fungus \emph{Physisporinus vitreus}, a pattern for which the mechanistic links deserve more study. 

Considering that the homes we studied differed greatly in their size, number of occupants, age, design, and much more, the fact that these variables influence so very little of fungal composition is striking. Houses, in general, favor some fungi relative to others and yet just which species appears to depend nearly exclusively on where the house is built.
 
\section{Discussion}\label{s:con}
In this paper, we introduced a nonparametric Bayesian model for identifying factors that influence microbiome composition, as well as a covariance estimator amenable to high-dimensional, binary data akin to that of \citet{Hall:2008}. The proposed model uses spike-and-slab variable selection to identify covariates that influence the occupancy probability of even a small subset of the taxa. It also utilizes a set of orthogonal, data-driven spatial basis functions and a Dirichlet process prior over their associated loadings to cluster the OTUs into groups of taxa that exhibit similar spatial responses, allowing dimension reduction in both the number of spatial locations and the number of taxa under consideration, greatly alleviating the computational burden compared to a parametric spatial model. 

We demonstrated via simulation study that the proposed model outperforms a na{\"i}ve nonspatial model and PERMANOVA in identifying influential covariates, and showed that violating the assumption of exchangeability of sampling locations underlying PERMANOVA leads to Type I error rates that are not well-controlled. We also showed that the proposed model is able to better identify low prevalence and/or small magnitude covariate effects as compared to a parametric spatial competitor.

We applied our proposed model to the indoor fungal microbiome from the Wild Life of Our Homes project as identified in \citet{Barberan:2015}. We were able to broadly substantiate their conclusion that geography and climate are the most influential factors affecting indoor fungal communities, and we provided additional detail in describing how factors affect particular taxa rather than simply classifying factors are influential or unimportant. 

This work primarily focused on the global hypothesis of whether or not a covariate influences microbiome composition as a whole.  However, the model also allows for local hypothesis tests of individual covariate values, which have not been fully explored here. We discussed the application and potential of these local tests, but did not rigorously test the true and false positive rates for covariate effects on individual taxa. An additional area of focus for future work is to expedite and improve the covariance estimation process to scale with large problems.

% Bibliography
\bibliographystyle{rss}
\bibliography{agents}

\section*{Appendix A: Model properties} \label{s:fullmodel}
With the assumptions from Section \ref{s:model}, the model for the latent process is
\begin{linenomath} \begin{align*} \label{eq:latent_sp}
Z_{j}(\bs) &= \beta_{0j} + \bX(\bs)\bbeta_{j} + \bPsi(\bs)\balpha_{j} + \epsilon_{j}(\bs) \nonumber \\
  &= \beta_{0j} + \sum\limits_{r=1}^{p}X_{r}(\bs)\delta_{jr}\gamma_{jr} + \sum\limits_{l=1}^{L} \psi_{l}(\bs)\alpha_{jl} + \epsilon_{j}(\bs).
\end{align*} \end{linenomath}

Conditionally on the cluster labels, $g_{j}$, the induced covariance between OTUs is
\beq \label{eq:condCross}
\mbox{Cov}\left[Z_j(\bs), Z_{j^{\prime}}(\bs^{\prime}) \, | \, g_{j}, g_{j^{\prime}}\right] = 
\begin{cases}
\; \rho\sum\limits_{l=1}^{L} \psi_{l}(\bs)\psi_{l}(\bs^{\prime}) & \mbox{if } g_{j} = g_{j^{\prime}} \\
\; 0 & \mbox{if } g_{j} \neq g_{j^{\prime}},
\end{cases} \nonumber
\eeq
for $j \neq j^{\prime}$, which is nonstationary in general. With $L$ sufficiently large, we can approximate any spatial covariance function by appealing to the $\KL$ theorem if the basis functions $\psi_{l}(\bs)$ are orthonormal (Karhunen, 1947). Marginally over the cluster labels, the induced covariance is 
\beq \label{eq:margCross}
\mbox{Cov}\left[Z_{j}(\bs), Z_{j^{\prime}}(\bs^{\prime})\right] = 2 \rho \varphi \left[\sum_{l=1}^{L} \psi_{l}(\bs)\psi_{l}(\bs^{\prime})\right], \nonumber
\eeq
where $j \neq j^{\prime}$. The probability that two OTUs are from the same cluster, $\varphi = \sum_{k=1}^{\infty} p_{k}^{2}$, controls the dependence between OTUs in this separable, spatial, multivariate covariance function. If $p_{k}$ is large for only a few clusters then $\varphi$ will be close to 1, and the OTUs will partition into a small number of clusters leading to strong dependence.  Otherwise, if the $p_{k}$ values are smaller and more uniform, indicating weaker groupings, then $\varphi$ will be close to 0. 

\begin{figure}[ht]
\centering
\includegraphics[width=0.65\textwidth]{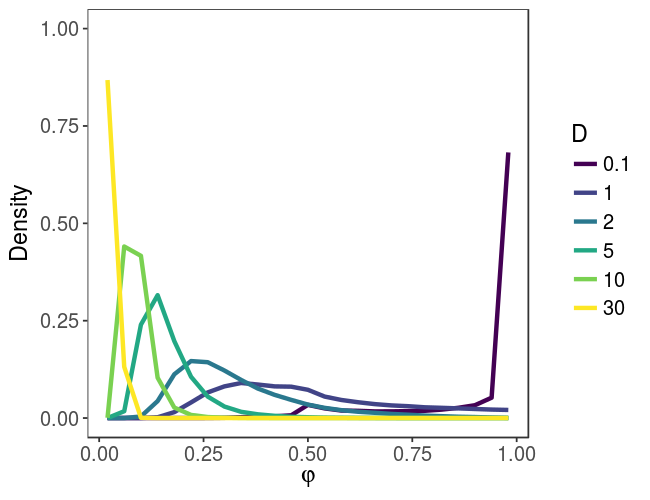}
\caption{Empirical distribution of $\varphi = \sum_{k=1}^{200}p_{k}^{2}$ for several values of the Dirichlet precision parameter, $D$.}
\label{f:sumProbsSq}
\end{figure}

Figure \ref{f:sumProbsSq} displays the empirical distribution of $\sum_{k=1}^{200} p_{k}^{2}$ for several values of the Dirichlet process precision parameter, $D$. 
For small values of $D$, the process favors fewer clusters reflected in $\sum_{k=1}^{200}p_{k}^{2}$ closer to 1. For large values of $D$, we see the reverse. We truncate the number of clusters at 200 for an example case of 1,000 taxa because $\mbox{E}[K \, | \, m, D] \approx D\mbox{log}\left[(m+D)/D\right]$ (Antoniak, 1974), where $K$ is the total number of groups created from $m$ taxa. Empirically, the median value for the maximum of $D$ based on its prior, discussed further in the following section, is 77 and thus a reasonable value for the maximum number of clusters is roughly 200.

\section*{Appendix B: Computing Details}
Recall that $i=1,\hdots,n$ indexes the sampling locations, $j=1,\hdots,m$ indexes the taxa, $r=1,\hdots,p$ indexes the covariates, $l=1,\hdots,L$ indexes the basis functions, and $k=1,\hdots,K$ indexes the clusters for the Dirichlet process, which are capped at $K=\mbox{min}(m, 500)$ for computational purposes. The full proposed model is
{\small 
\begin{flalign*}
\; & Y_{j}(\bs_{i}) \, | \, Z_{j}(\bs_{i}) = \mathbbm{1}\{Z_{j}(\bs_{i}) > 0 \} & \\
& Z_{j}(\bs_{i}) \, | \, \bX, \beta_{0j}, \bbeta_{j}, \balpha_{j}, \rho \indep \mbox{N}\left(\beta_{0j} + \bX(\bs_{i})\bbeta_{j} + \bPsi(\bs_{i})\balpha_{j}, 1-\rho \right) & \\
& \balpha_{j} \, | \, g_{j} = k, \bmu_{1}, \hdots, \bmu_{K} = \bmu_{k} & \\
& \bmu_{k} \, | \, \bmu_{0}, \rho \iid \mbox{N}_{L}\left(\bmu_{0}, \rho\bm{\mbox{I}}_{L}\right) & \\
& \bmu_{0} \, | \, \tau_{\mu_{0}} \sim \mbox{N}_{L}\left(\bm{0}, \tau^{-1}_{\mu_{0}}\bm{\mbox{I}}_{L}\right), \, \tau_{\mu_{0}} \sim \mathrm{Gamma}(a_{\mu_{0}}=0.1, b_{\mu_{0}}=0.1) & \\
& \rho \sim \mathrm{U}(0, 1) & \\
& \mathrm{P}(g_{j}=k) = p_{k} = V_{k}\prod\limits_{u < k}(1-V_{u}) \text{ for } k>1 \text{ and } p_{1} = V_{1}& \\
& V_{u} \, | \, D \iid \mathrm{Beta}(1, D) \text{ for } u=1, \hdots, K-1 \text{ and } V_{K}=1 & \\
& D \sim \text{Gamma}(a_{d}=0.1, b_{d}=0.1) & \\
& \beta_{0j} \, | \, \tau_{0} \iid \mbox{N}(0, \tau^{-1}_{0}), \, \tau_{0} \sim \mathrm{Gamma}(a_{0}=0.1, b_{0}=0.1) & \\ 
& \beta_{jr} \, | \, \delta_{jr}, \gamma_{jr} = \delta_{jr}\gamma_{jr} & \\
& \delta_{jr} \, | \, \pi_{r} \indep \mathrm{Bernoulli}(\pi_{r}) & \\
& \mbox{P}(\pi_{r} \, | \, \omega, \theta) = \omega\left[\frac{1}{\mathrm{B}(1, \theta)}(1-\pi_{r})^{\theta-1}\right] + (1 - \omega) \text{ for } \omega \in [0, 1] \text{ and } \theta \ge 1 \text{, fixed} & \\
& \gamma_{jr} \, | \, \tau_{r} \indep \mbox{N}(0, \tau_{r}^{-1}), \, \tau_{r} \iid \mathrm{Gamma}(a_{r}=1, b_{r}=2.7) &
\end{flalign*}
}

We have chosen to follow the approach of allowing $a_{d}, b_{d} \rightarrow 0$ by setting them to small values (Escobar and West, 1995; Navarro et al., 2006). Recently, alternative approaches have been developed that attempt to correct for pitfalls wherein learning about $D$ is difficult and therefore inference is sensitive to its prior specification in small sample problems (Dorazio, 2009; Murugiah and Sweeting, 2012). However, these approaches are not feasible for high-dimensional problems because of a reliance on unsigned Stirling numbers of the first kind in Dorazio (2009) or on an extensive performance study in Murugiah and Sweeting (2012), which are computationally demanding, if not impossible. 

The remaining hyperprior parameters are chosen as the standard uninformative values, with the exception of $a_{r}$ and $b_{r}$. In this setting, the usual values of $0.1$ caused numerical instability within matrix inversions.  To resolve this, the values $a_{r}=1$ and $b_{r}=2.7$ were chosen to closely match the $\mbox{Gamma}(0.1, 0.1)$ distribution while restricting the maximum induced variance slightly to improve computational stability. An alternative solution is to assume that the variance of the magnitudes is the same for all covariates in the prior, i.e., $\tau_{1} = \hdots = \tau_{r} = \tau$, and use the standard prior $\tau \sim \mbox{Gamma}(0.1, 0.1)$. Both options correct the numerical instability and based on simulation testing, selecting either option is effective for variable selection.  Thus, unless there is outside expertise to suggest otherwise, we recommend the parsimonious option and suggest fitting the models using a common variance, where $\gamma_{jr} \, | \, \tau \iid \mbox{N}(0, \tau^{-1})$, and $\tau \sim \mbox{Gamma}(a_{\gamma}=0.1, b_{\gamma}=0.1)$. In the implementation details to follow, we give the update for the more complex case, but it should be stated that we use the common variance option as the default value in our implementation and that we utilized this simplification in all of the results presented in the body of the paper.

Posterior samples are drawn using Markov chain Monte Carlo (MCMC) with convergence monitored by inspecting trace plots.  Most model parameters can be updated via Gibbs sampling, with the exception of the variance of the structural component of the residual dependence, $\rho$, which is updated using the Metropolis algorithm.

\subsection*{Metropolis sampling for $\rho$}
The log posterior distribution for $\rho$, conditional on all other parameters, is given by
{\small 
\begin{flalign*}
\; \ell(\rho \, | \, \cdots) \propto &- \frac{nm}{2}\mbox{log}(1-\rho) - \frac{LK}{2}\mbox{log}(\rho) - \frac{1}{2\rho}\sum_{k=1}^{K}\big\|\bmu_{k}-\bmu_{0}\big\|^{2}_{2} & \\
& - \frac{1}{2(1-\rho)}\sum\limits_{j=1}^{m}\big\|\bZ_{j} - \beta_{0j}\bm{1}_{n} - \bX\bbeta_{j} - \bPsi\balpha_{j}\big\|^{2}_{2}. &
\end{flalign*}
}%
Thus, $\rho$ cannot be updated using Gibbs sampling and instead requires a Metropolis update. Because $\rho$ is bound by the interval $[0, 1]$, we use the logit transformation and work with the continuous variable $\mbox{logit}(\rho) = \mbox{log}\left(\frac{\rho}{1-\rho}\right)$. At each iteration, we propose a candidate, $\mbox{logit}(\rho^{*}) \sim \mbox{N}\{\mbox{logit}(\rho), \sigma^{2}_{M}\}$, where $\sigma^{2}_{M}$ is adapted within the burn-in period to maintain an acceptance rate $\in [0.3, 0.7]$.

\subsection*{Gibbs sampling}
All other model parameters are drawn from their full conditional distributions using Gibbs sampling.  The full conditional distributions are as follows:
{\small 
\begin{linenomath} \begin{flalign*}
\; & Z_{j}(\bs_{i}) \, | \cdots \sim \mbox{TN}\{\beta_{0j} + \bX(\bs_{i})\bbeta_{j} + \bPsi(\bs_{i})\balpha_{j}, (1-\rho); (l_{ij}, u_{ij})\}, & \\
 & \qquad \text{where } \mbox{TN}\{\mu, \sigma^{2}; (a, b)\} \text{ denotes the } \mbox{N}(\mu, \sigma^{2}) \text{ distribution truncated to } & \\
 & \qquad \text{lie in the interval } (a, b), \text{ and } (l_{ij}, u_{ij}) = (0, \infty) \text{ if } Y_{j}(\bs_{i}) = 1 \text{ or }& \\
 & \qquad (l_{ij}, u_{ij}) = (-\infty, 0) \text{ if } Y_{j}(\bs_{i}) = 0, & \\
& \beta_{0j} \, | \cdots \sim \mbox{N}\left(\frac{1}{n+(1-\rho)\tau_{0}}\sum_{i=1}^{n}\left[Z_{j}(\bs_{i}) - \bX(\bs_{i})\bbeta_{j} - \bPsi(\bs_{i})\balpha_{j}\right], \frac{1-\rho}{n + (1-\rho)\tau_{0}}\right), & \\
& \tau_{0} \, | \cdots \sim \mbox{Gamma}\left(a_{0} + \frac{m}{2}, b_{0} + \frac{1}{2}\sum_{j=1}^{m}\beta_{0j}^{2} \right), & \\
& \mbox{P}\left(g_{j}=k \, | \cdots \right) = \frac{p_{k}\mbox{P}(\bZ_{j} \, | \, \balpha_{j}=\bmu_{k})}{\sum\limits_{c=1}^{K}p_{c}\mbox{P}(\bZ_{j} \, | \, \balpha_{j}=\bmu_{c})}, & \\
& V_{u} \, | \cdots \sim \mbox{Beta}(1 + n_{k}, D + n_{>k}) \text{ for } u=1,\hdots,K-1 & \\
 & \qquad \text{where } n_{k} = \sum_{j=1}^{m} \mathbb{I}\{g_{j} = k\} \text{ is the number of OTUs in cluster } k & \\
 & \qquad \text{and } n_{>k} = \sum_{j=1}^{m} \mathbb{I}\{g_{j} > k\} \text{ is the number of OTUs in clusters above } k, & \\
& D \, | \cdots \sim \mbox{Gamma}\left(a_{d}+K-1 , b_{d} - \sum\limits_{u=1}^{K-1}\mbox{log}(1-V_{u}) \right), & \\
& \bmu_{k} \, | \cdots \sim \mbox{N}_{L}\Bigg(\left[\frac{n_{k}}{1-\rho}\bPsi^{\prime}\bPsi + \frac{1}{\rho}\right]^{-1}\left[\frac{1}{\rho}\bmu_{0} + \frac{1}{1-\rho}\bPsi^{\prime}\sum\limits_{j: g_{j}=k}\bZ_{j} - \beta_{0j}\bm{1}_{n} - \bX\bbeta_{j}\right], & \\
& \qquad \qquad \qquad \; \left[\frac{n_{k}}{1-\rho}\bPsi^{\prime}\bPsi + \frac{1}{\rho}\right]^{-1}\Bigg), & \\
& \qquad \text{ if } n_{k} > 0, \text{ otherwise } \bmu_{k} \text{ is drawn from the prior distribution}, & \\
& \bmu_{0} \, | \cdots \sim \mbox{N}_{L}\left(\frac{1}{K + \rho\tau_{\mu_{0}}}\sum\limits_{k=1}^{K}\bmu_{k}, \frac{\rho}{K + \rho\tau_{\mu_{0}}}\right), & \\
& \tau_{\mu_{0}} \, | \cdots \sim \mbox{Gamma}\left(a_{\mu_{0}} + \frac{L}{2}, b_{\mu_{0}} + \frac{1}{2}\bmu_{0}^{\prime}\bmu_{0}\right), & \\
& \bm{\gamma}_{j} \, | \cdots \sim \mbox{N}_{p}\Bigg(\left[\frac{1}{1-\rho}\bX_{*j}^{\prime}\bX_{*j} + \bT_{\gamma}\right]^{-1}\frac{1}{1-\rho}\bX_{*j}^{\prime}\left[\bZ_{j} - \beta_{0j}\bm{1}_{n} - \bPsi\balpha_{j}\right], & \\
& \qquad \qquad \qquad \; \left[\frac{1}{1-\rho}\bX_{*j}^{\prime}\bX_{*j} + \bT_{\gamma}\right]^{-1}\Bigg) & \\
 & \qquad \text{ where } \bT_{\gamma} = \mbox{diag}\{\tau_{1}, \hdots, \tau_{p}\}, \, \bm{\Lambda}_{j} = \mbox{diag}\{\delta_{j1}, \hdots, \delta_{jp}\}, \text{ and } \bX_{*j} = \bX\bm{\Lambda}_{j}, & \\
& \tau_{r} \, | \cdots \sim \mbox{Gamma}\left(a_{r} + \frac{m}{2}, b_{r} + \frac{1}{2}\sum\limits_{j=1}^{m}\gamma_{jr}^{2}\right), & \\
& \delta_{jr} \, | \cdots \sim \mbox{Bernoulli}\left(\pi^{*}_{jr}\right) & \\
 & \qquad \text{ where } \mbox{logit}(\pi^{*}_{jr}) = \mbox{log}(\pi_{r}) - \mbox{log}(1 - \pi_{r}) & \\
& \qquad \qquad \qquad \qquad \qquad - \frac{1}{2(1-\rho)}\sum\limits_{i=1}^{n}\Bigg[Z_{j}(\bs_{i}) - \beta_{0j} - \bPsi(\bs_{i})\balpha_{j} & \\
& \qquad \qquad \qquad \qquad \qquad \qquad \qquad \qquad \qquad - \sum_{q \neq r}X_{q}(\bs_{i})\delta_{jq}\gamma_{jq}- X_{r}(\bs_{i})\gamma_{jr}\Bigg]^{2} & \\
 & \qquad \qquad \qquad \qquad \qquad + \frac{1}{2(1-\rho)}\sum\limits_{i=1}^{n}\Bigg[Z_{j}(\bs_{i}) - \beta_{0j} - \bPsi(\bs_{i})\balpha_{j} & \\
& \qquad \qquad \qquad \qquad \qquad \qquad \qquad \qquad \qquad - \sum_{q \neq r}X_{q}(\bs_{i})\delta_{jq}\gamma_{jq}\Bigg]^{2}, & \\
& \mbox{P}(\pi_{r} \, | \, \cdots) = W_{r}\mbox{Beta}(1+M_{r}, \theta+m-M_{r}) + (1-W_{r})\mbox{Beta}(1+M_{r}, 1+m-M_{r}) & \\
 & \qquad \text{where } \mbox{B}(a, b) \text{ is the Beta function, } M_{r} = \sum\limits_{j=1}^{m}\delta_{jr}, & \\
 & \qquad \text{ and } W_{r} = \frac{\omega\theta\mbox{B}(1+M_{r}, \theta+m-M_{r})}{\omega\theta\mbox{B}(1+M_{r}, \theta+m-M_{r}) + (1-\omega)\mbox{B}(1+M_{r}, 1+m-M_{r})}. &
\end{flalign*} \end{linenomath}
} 

\end{document}